\documentclass[twocolumn,showpacs,preprintnumbers,amsmath,amssymb]{revtex4}
\usepackage{graphicx}

\begin{document}

\title{Input-output theory for fermions in an atom cavity}
\author{C. P. Search, S. P\"{o}tting, W. Zhang, and P. Meystre}
\affiliation{Optical Sciences Center, The University of Arizona,
Tucson, AZ 85721}
\date{\today}

\begin{abstract}
We generalize the quantum optical input-output theory developed
for optical cavities to ultracold fermionic atoms confined in a
trapping potential, which forms an "atom cavity". In order to
account for the Pauli exclusion principle, quantum Langevin
equations for all cavity modes are derived. The dissipative part
of these multi-mode Langevin equations includes a coupling between
cavity modes. We also derive a set of boundary conditions for the
Fermi field that relate the output fields to the input fields and
the field radiated by the cavity. Starting from a constant uniform
current of fermions incident on one side of the cavity, we use the
boundary conditions to calculate the occupation numbers and
current density for the fermions that are reflected and
transmitted by the cavity.
\end{abstract}
 \maketitle

\section{Introduction}
In light of the remarkable achievement of Bose-Einstein
condensation in 1995 \cite{BEC}, there has been a growing
application of ideas from quantum optics to matter waves. This new
field of atom optics \cite{atomoptics} has included both
theoretical and experimental investigations of matter wave
coherence \cite{goldstein,andrews,miesner,burt,kohl}, atom lasers
\cite{alasertheory,alaserexper}, non-linear effects in matter
waves including matter wave mixing \cite{nwmtheory,nwmexper},
parametric amplification and squeezing in coupled optical and
matter waves \cite{moore}. However the extension of these ideas to
degenerate Fermi gases has proven difficult because of the Pauli
exclusion principle, which prohibits one from developing simple
theoretical models based on only a few normal modes of the
Schr\"{o}dinger field. In addition to this, Fermi fields do not
possess a classical limit analogous to the coherent state for Bose
fields thereby making it impossible to develop semiclassical
mean-field theories such as the Gross-Pitaevskii equation for Bose
fields.

All in all, the physical intuition obtained from quantum optics
cannot be directly applied to theoretical investigations of
fermions. It therefore seems necessary that in order to make
progress in the theory of fermionic atom optics, fundamental model
systems in quantum optics need to be reanalyzed from first
principles. Recent work in this direction indicates that four-wave
mixing and coherent amplification of matter waves can occur in
fermionic systems as a result of cooperative many-particle quantum
interference analogous to Dicke superradiance
\cite{moore2,ketterle,villian}.

The purpose of this paper is to consider another model system, the
atomic analog of an optical cavity with two partially transmittive
mirrors. A schematic of our system is illustrated in Fig. 1. It
consists of an atom cavity formed by two potential barriers with a
finite number of bound states. The cavity states are coupled to a
continuum of free particle states on either side of the cavity via
tunnelling through the barriers. Our goal is to develop an
input-output theory for fermions in the atom cavity that allows
one to calculate the field radiated out of the cavity in terms of
the field incident on the cavity.

The input-output theory for a single mode of a lossy optical
cavity was developed by Collett and Gardiner in the form of
quantum Langevin equations for the cavity mode \cite{collett}. The
great utility of this theory is that it allows one to incorporate
the effects of quantum noise on the output field transmitted by
the cavity as well as in the intracavity dynamics. Collett and
Gardiner's formalism has been extended to bosonic matter fields in
order to model the output coupling of atoms from a Bose-Einstein
condensate in a single mode of an atom trap \cite{hope}. As we
will show below, the necessity of treating all modes of the atom
cavity for fermions leads to novel features not present in the
single mode bosonic theories. Most significantly, the eigenstates
of the cavity become coupled due to their mutual interaction with
the same external continuum states. Secondly, the coupling of the
reservoir modes to all cavity modes leads to the creation of
coherences between fermions occupying different single particle
modes in the radiated field even if the incident field is
completely incoherent.

The outline of the paper is as follows: In section II, we present
our physical model for an atom cavity coupled to a continuum of
reservoir states. In section III, we derive a set of quantum
Langevin equations for the fermionic annihilation and creation
operators of the eigenstates of the cavity in terms of both the
input and output fields. These results are generalized to a two
sided cavity in section IV. In section V, we consider a constant
current of fermions incident on one side of the atom cavity and
calculate the steady state statistics of the fermions transmitted
through the other side of the cavity. In appendix A, we show that
in a manner similar to the bosonic case, the presence of noise
operators in the Langevin equations is necessary to preserve the
anticommutation relations for the fermion operators. In appendix B
we derive explicit forms for the coupling constants, which connect
the intracavity modes to external continuum modes via a tunnelling
Hamiltonian.

\begin{figure}
\includegraphics*[width=0.9\columnwidth]{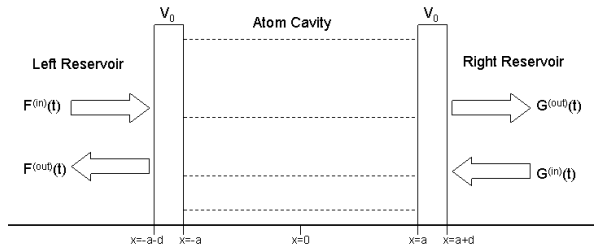}
\caption{Schematic diagram of 1D atom cavity system. }\label{fig1}
\end{figure}

\section{Physical Model}
A physical schematic of our system is illustrated in Fig. 1. For
simplicity, we restrict ourselves to one spatial dimension, which
allows space to be divided up into five distinct regions. The
region $-a\leq x \leq a$ between the two potential barriers of
height $V_{0}$ represents the atom cavity. For thick barriers, the
number of bound states of the cavity, $N+1$, is given by
$N\pi/2<\beta=\sqrt{2mV_{0}a^2/\hbar^2}\leq (N+1)\pi/2$ where $m$
is the mass of the atoms \cite{merzbacher}. We will focus on the
case where $N \gg 1$. The regions $-L-a-d<x<-a-d$ and
$a+d<x<L+a+d$ represent the left and right reservoirs,
respectively. $L$ is the length of the reservoir region. We let it
go to infinity so that the fermions are described in terms of a
continuum of free particle plane wave states. Atoms located in the
reservoir regions with energies less the $V_{0}$ couple to the
cavity states by tunnelling through the potential barriers located
at $-a-d\leq x < -a$ and $a< x \leq a+d$.

Since the wave functions for atoms with energies greater than
$V_{0}$ are not spatially localized in either the cavity or the
reservoir regions, we restrict ourselves to single particle states
with energies less than $V_{0}$. In this case we can meaningfully
speak of left/right reservoir states and cavity states since the
single particle wave functions decay exponentially inside the
potential barriers. Such a restriction is valid provided the
initial state does not contain any occupied states with energies
greater than $V_0$ and two-body collisional interactions between
atoms, which can cause atoms to be scattered into higher energy
states, are negligible. The latter condition will indeed be
satisfied for ultra-cold spin polarized fermions since $s$-wave
collisions are forbidden and $p$-wave collisions are negligible at
these temperatures. Under these conditions, the states with
energies larger than $V_0$ are not coupled to states with energies
less than $V_0$.

The second quantized Hamiltonian for the cavity-reservoir system
in the subspace of states with energies below the barrier is
\begin{equation}
H=H_S+H_L+H_R+H_{SR}+H_{SL} \label{HTOT}
\end{equation}
where $H_S$, $H_L$, and $H_R$ are the free Hamiltonians for the
system (i. e. the atom cavity) and the left and right reservoir
states, respectively,
\begin{eqnarray}
H_S=\sum_{n=0}^N\hbar\Omega_n c^{\dagger}_{n}c_{n} \\
H_L=\sum_{k}\hbar\omega_k a^{\dagger}_k a_k \\
H_R=\sum_{k}\hbar\omega_k b^{\dagger}_k b_k.
\end{eqnarray}
Here, $c_n$ is a fermionic annihilation operator that destroys an
atom in the cavity with wave function $\phi^{(s)}_n(x)$ and energy
$\hbar\Omega_n=\hbar^2K_n^2/2m$. Similarly, $a_k$ and $b_k$ are
fermionic annihilation operators that destroy an atom in the left
and right reservoirs, respectively, with the wave function
\[
\phi^{(l,r)}_k(x)=\exp(ik(x \pm (a+d))/L^{1/2}
\]
and energy $\hbar\omega_k=\hbar^2k^2/2m$ in the regions outside
the barrier.

The coupling between the system and the reservoirs, $H_{SR}$ and
$H_{SL}$, is given by effective tunnelling Hamiltonians
\cite{cohen,prange,bardeen,meier}
\begin{eqnarray}
H_{SL}=i\hbar\sum_{n,k}\left[\kappa_{n,k}c^{\dagger}_n a_k -
\kappa_{n,k}^{*}a^{\dagger}_k c_n \right] \\
H_{SR}=i\hbar\sum_{n,k}\left[\tilde{\kappa}_{n,k}c^{\dagger}_n b_k
- \tilde{\kappa}_{n,k}^{*}b^{\dagger}_k c_n \right].
\end{eqnarray}
In all cases the summation is restricted to those states with
energies below the barrier. Explicit expressions for the
tunnelling matrix elements, $\kappa_{n,k}$ and
$\tilde{\kappa}_{n,k}$, are given in Appendix B. We note here that
in 1-D, the coupling constants depend only on the magnitude of $k$
and not its sign.

In contrast to quantum optical systems, which are often
approximated as a single cavity mode with a large occupation
number, a full multi-mode treatment is required for fermions even
if the number of fermions in the cavity is small ($\sim 1$). This
is because of the Pauli principle, which forbids more than one
atom from occupying the same cavity state, and thereby prevents
one from singling out a particular state as being more important
than the rest.

We conclude this section by noting that the general results
presented below do not depend on the precise nature of our
physical model. We use a stepwise constant potential because it
leads to simple analytic results for the coupling between the
reservoirs and cavity states. Our model system in Eq. (\ref{HTOT})
could be applied to any fermion system in which a finite number of
discrete states are linearly coupled to a dense continuum of
states.

\section{single sided cavity}
We now derive a set of integro-differential equations of motion
for the cavity operators that only involve the initial or final
state of the reservoir operators. We proceed by formally
integrating the equations of motion for the reservoir operators
and substituting these back into the equations for the cavity
operators. In this section, we set $H_R=H_{SR}=0$ so that the atom
cavity is coupled to a single reservoir. This is analogous to an
optical cavity with a single partially transmittive mirror. The
inclusion of the right reservoir is summarized in the following
section.

It is convenient to work with slowly varying operators in the
interaction representation,
\begin{equation}
c_n(t)=e^{-i\Omega_nt}\hat{c}_n(t), \: a_k(t)=e^{-i\omega_k
t}\hat{a}_k(t).
\end{equation}
By formally integrating the Heisenberg equations of motion for
$\hat{a}_k(t)$ from the initial time $t_0$ to $t$,
\begin{equation}
\hat{a}_k(t)=\hat{a}_k(t_0)-\sum_m \kappa^{*}_{m,k} \int_{t_0}^t
dt' e^{i(\omega_k-\Omega_m)t'} \hat{c}_m(t')
\end{equation}
where $\hat{a}_k(t_0)$ are the operators for the input field
incident on the cavity barrier, and substituting this solution
into the equations of motion for $\hat{c}_n(t)$, we obtain
\begin{widetext}
\begin{equation}
\frac{d}{dt}\hat{c}_n(t)=\sum_{k}\kappa_{n,k}e^{-i(\omega_k-\Omega_n)t}\hat{a}_k(t_0)-
\sum_m e^{i(\Omega_n-\Omega_m)t} \int_{0}^{t-t_0}d\tau
\alpha_{n,m}(\tau) \hat{c}_m(t-\tau). \label{lang1}
\end{equation}
\end{widetext}
Here,  $\alpha_{n,m}(\tau)$ is a reservoir correlation function,
\begin{equation}
\alpha_{n,m}(\tau)=\sum_{k}\kappa_{n,k}\kappa^{*}_{m,k}e^{i(\Omega_m-\omega_k)\tau}.
\end{equation}
that decays to zero in a characteristic time $\tau_c$ due to the
destructive interference between the different oscillations. Note
that $\tau_c$ depends, in general, on the cavity states $n$ and
$m$ that are coupled by $\alpha_{n,m}(\tau)$, and $\tau_c^{-1}$ is
on the order of the bandwidth of the reservoir, $V_0/\hbar$.
Furthermore, if we assume that $\hat{c}_m(t)$ only changes
significantly over a time scale $T_m \gg \tau_c$, then we can make
the Markov approximation by setting
$\hat{c}_m(t-\tau)=\hat{c}_m(t)$ in Eq. (\ref{lang1}). For time
intervals $t-t_0 \gg \tau_c$, we can then make the replacement
\begin{equation}
\int^{t-t_0}_0 d\tau \alpha_{n,m}(\tau)\approx \int^{\infty}_0
d\tau \alpha_{n,m}(\tau) \label{integrate}
\end{equation}
where the later expression is given by
\begin{equation}
\int^{\infty}_0 d\tau
\alpha_{n,m}(\tau)=\gamma_{n,m}+i\Delta_{n,m},
\end{equation}
with
\begin{equation}
\gamma_{n,m}=\pi \sum_k
\kappa_{n,k}\kappa^{*}_{m,k}\delta(\Omega_m-\omega_k)
\label{damp1}
\end{equation}
and
\begin{equation}
 \Delta_{n,m}=P\sum_k
\kappa_{n,k}\kappa^{*}_{m,k}\frac{1}{\Omega_m-\omega_k}.
\end{equation}
These expressions are defined in the continuum limit that $\sum_k
\rightarrow (L/2\pi)\int dk$.

The Markov approximation assumes that the correlation function
decays very rapidly, which requires that $\kappa_{n,k}$ vary
slowly with $k$. Fig. 2, shows the correlation function
$|\alpha_{n,n}(\tau)|$ for the highest energy bound state in the
potential well. This state has the longest correlation time since
the $|\kappa_{n,k}|^2$ are the largest for this state and because
the summation over reservoir states is restricted to states with
energies less than $V_0$. For the three cases plotted in Fig. 2,
the smallest value of $\gamma_{n,n}^{-1}$ among all the cavity
states is $0.55\Omega_0^{-1}$ ($d/a=0.0001$), $0.62\Omega_0^{-1}$
($d/a=0.001$), and $1.36\Omega_0^{-1}$ ($d/a=0.01$). In each case
we see that the correlation function goes to zero in a time much
shorter than $T_n \approx \gamma_{n,n}^{-1}$. This shows that the
Markov approximation is a very good approximation for our system
with $T_n/\tau_C >10^2$.

\begin{figure}
\includegraphics*[width=0.9\columnwidth]{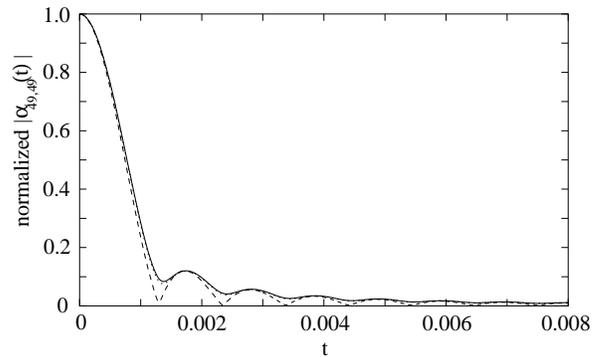}
\caption{Plot of $\alpha_{n,n}(\tau)$ for $n=49$ and
$d/a=0.0001$(solid line), $0.001$(dotted line), $0.01$ (dashed
line). Times are measured in units of $\Omega_0^{-1}$. $V_0$ and
$a$ were chosen so that the cavity contains $50$ bound states.}
\end{figure}

By combining the above results, we obtain a quantum Langevin
equation for each of the cavity modes,
\begin{equation}
\dot{c}_n(t)=-i\Omega_nc_n(t)-\sum_m
(\gamma_{n,m}+i\Delta_{n,m})c_m(t)+F^{(in)}_n(t). \label{lang_in}
\end{equation}

In Eq. (\ref{lang_in}), we have defined the {\it input noise
operator} $F^{(in)}_n(t)$ as
\begin{equation}
F^{(in)}_n(t)=\sum_k\kappa_{n,k}e^{-i\omega_k(t-t_0)}a_k(t_0)\equiv
\sum_k\kappa_{n,k}a^{(in)}_k(t)
\end{equation}
where
\[
a^{(in)}_k(t)=a_k(t_0)\exp (-i\omega_k(t-t_0))
\]
is the annihilation operator for mode $\phi^{(l)}_k(x)$ of the
input Fermi field at time $t$, and the total input field operator
is therefore
\[
\Psi^{(in)}(x,t)=\sum_k a_k^{(in)}(t)\phi^{(l)}_k(x).
\]
That is, $\Psi^{(in)}(x,t)$ is the free Fermi field that
propagates from the initial time $t_0$ to $t$ in the Heisenberg
picture.

Since the initial reservoir operators obey the anticommutation
relations $\{a_k(t_0),a^{\dagger}_{k'}(t_0)\}=\delta_{k,k'}$ and
$\{a_k(t_0),a_{k'}(t_0)\}=0$, it is easy to show that the noise
operators obey the anticommutation relations,
\begin{eqnarray}
\{F^{(in)}_n(t),F^{(in)\dagger}_m(t-\tau)\}=e^{-i\Omega_m\tau}\alpha_{n,m}(\tau)
\label{acn_in}
\\ \{F^{(in)}_n(t),F^{(in)}_m(t-\tau)\}=0 \label{acn_in2}
\end{eqnarray}
Furthermore, if we restrict ourselves to time scales much longer
than the correlation time $\tau_c$, then we can approximate the
correlation function in Eq. (\ref{acn_in}) by a delta function
times the area under $\alpha_{n,m}(\tau)$, so that
\begin{equation}
\{F^{(in)}_n(t),F^{(in)\dagger}_m(t-\tau)\}\approx
2\gamma_{n,m}\delta(\tau) \label{noisecomm}.
\end{equation}

Before proceeding, there are several features of Eq.
(\ref{lang_in}) that are worth mentioning. First, the dissipative
term $\sum_m (\gamma_{n,m}+i\Delta_{n,m})c_m(t)$ gives a damping
term plus an energy shift for $n=m$ while for $n\neq m$ there is a
non-zero coupling between cavity states. The coupling between
cavity states is a result of all states coupling to the same
reservoir, which leads to an indirect coupling between cavity
states. Second, the noise operators couple all of the reservoir
states to each of the cavity states. Moreover, the noise operators
are different for each cavity state due to the $n$-dependence of
the coupling constants.

Instead of solving for the reservoir operators in terms of the
initial time $t_0$, one can instead solve for $\hat{a}_k(t)$ in
terms of a final time $t_1>t$,
\begin{equation}
\hat{a}_k(t)=\hat{a}_k(t_1)+\sum_m \kappa^{*}_{m,k} \int_{t}^{t_1}
dt' e^{i(\omega_k-\Omega_m)t'} \hat{c}_m(t'). \label{a_fin}
\end{equation}
The operators $\hat{a}_k(t_1)$ represent the modes of the output
field that contain the field radiated by the cavity at earlier
times. By substituting this expression into the equations of
motion for $\hat{c}_n(t)$, making the Markov approximation in the
integrand, and transforming back to the Heisenberg representation
one obtains
\begin{equation}
\dot{c}_n(t)=-i\Omega_nc_n(t)+\sum_m
(\gamma_{n,m}-i\Delta_{n,m})c_m(t)+F^{(out)}_n(t).
\label{lang_out}
\end{equation}
Here, $F^{(out)}_n(t)$ is the {\it output field noise operator}
for state $n$,
\begin{equation}
F^{(out)}_n(t)=\sum_k\kappa_{n,k}e^{-i\omega_k(t-t_1)}a_k(t_1)\equiv
\sum_k\kappa_{n,k}a^{(out)}_k(t)
\end{equation}
where we have defined the output field annihilation operator
$a^{(out)}_k(t)$ for the mode $\phi^{(l)}_k(x)$. The
$a^{(out)}_k(t)$ are related to the output field of the reservoir
by,
\[
\Psi^{(out)}(x,t)=\sum_k a_k^{(out)}(t)\phi^{(l)}_k(x).
\]
It is clear that $\Psi^{(out)}(x,t)$ represents the free Fermi
field for the reservoir that propagates from $t$ to the final time
$t_1$. It is easy to see that if the $a_k(t_1)$ obey normal
fermionic anticommutation relations, then the anticommutators for
the output noise operators are the same as Eqs.
(\ref{acn_in}-\ref{acn_in2}).

The boundary condition for the barrier separating the cavity from
the reservoir is obtained by subtracting  Eq. (\ref{lang_in}) from
Eq. (\ref{lang_out}),
\begin{equation}
F^{(in)}_n(t)-F^{(out)}_n(t)=2\sum_m \gamma_{n,m}c_m(t) \label{bc}
\end{equation}
Equation (\ref{bc}) relates the noise operator for the output
field to the input noise operator reflected by the barrier and the
field radiated by the cavity. It is of the same form as the
boundary condition for an optical cavity \cite{collett} except
that in our case there are separate noise operators for each
cavity modes since we cannot, in general, assume that the coupling
constants are independent of the cavity state. In Appendix A, we
use Eq. (\ref{bc}) and (\ref{lang_in}) to derive the
anticommutation relations between the cavity operators and the
noise operators at arbitrary times.

Equation (\ref{bc}) is not very useful since it is the mode
operators of the output field, $a_k^{(out)}(t)$, that are needed
to calculate properties of the output field such as mode
occupation statistics, current density, etc. Therefore, we must
extract from Eq. (\ref{bc}) a boundary condition for the
annihilation operators of the modes of the input and output
fields.

First we note that in the limit of an infinite potential barrier
the system-reservoir coupling vanishes, $\kappa_{n,k}\equiv 0$. In
this limit, a fermion incident from the left is perfectly
reflected by the barrier. Since $\Psi^{(in)}(x,t)$ and
$\Psi^{(out)}(x,t)$ are the free Schr\"{o}dinger fields that
propagate forward in time from $t_0\rightarrow -\infty$ to $t$ and
from $t$ to $t_1\rightarrow \infty$, respectively, it follows that
\begin{equation}
a_k(t_0)e^{i\omega_k t_0}=-a_{-k}(t_1)e^{i\omega_k t_1}.
\label{inoutbc}
\end{equation}
The total field is the incident plus reflected fields,
\begin{equation}
\Psi(x,t)=\Psi^{(in)}(x,t)+\Psi^{(out)}(x,t). \label{total}
\end{equation}
It follows from (\ref{inoutbc}) and (\ref{total}) that
$\Psi(-a-d,t)=0$ and that the eigenmodes of the reservoir are
standing waves for $V_0 \rightarrow \infty$. Note that
(\ref{inoutbc}) and (\ref{total}) are second quantized versions of
the relations for the incident and reflected wave functions from
an infinite potential barrier \cite{merzbacher}.

For the reservoir, the displacement operator is given by
\[
D(x)=e^{-iPx/\hbar}
\]
where $P=\sum_k \hbar k a^{\dagger}_k a_k$ is the momentum
operator for the reservoir \cite{schrodrep}. Multiplying
(\ref{bc}) by $D(x)$ on the left and $D^{\dagger}(x)$ on the right
gives after some manipulation \cite{manipnote},
\begin{eqnarray}
\sum_k\kappa_{n,k}\left(a_k^{(in)}(t)-a_{-k}^{(out)}(t)\right)e^{ikx}
\nonumber \\
=2\pi\sum_{m,k}\kappa_{n,k}\kappa_{m,k}^{*}\delta(\Omega_m-\omega_k)c_m(t)e^{ikx}.
\label{bcproj}
\end{eqnarray}
Multiplying Eq. (\ref{bcproj}) by
$\int_{-L-a-d}^{-a-d}dxe^{-ik'x}$ and taking the limit
$L\rightarrow \infty$ gives the boundary condition for the modes
of the input and output fields,
\begin{equation}
a_k^{(in)}(t)-a_{-k}^{(out)}(t)=2\pi\sum_m\kappa_{m,k}^*\delta(\Omega_m-\omega_k)c_m(t)
\label{bcmode}
\end{equation}
Physically, Eq. (\ref{bcmode}) says that the difference between
the mode of the output field propagating away from the barrier
with momentum $-\hbar k$ is the input field with momentum $\hbar
k$ reflected by the barrier plus the field radiated by the cavity.
Furthermore, only those states of the cavity that have the same
energy as the reservoir mode can radiate into that mode. For a one
dimensional system, there will only be a single cavity mode that
contributes to the right hand side of Eq. (\ref{bcmode}). The
$\delta(\Omega_m-\omega_k)$ comes from Eq. (\ref{integrate}) where
we assumed times much longer than width of the reservoir
correlation function. Hence, it follows from the uncertainty
relation $\Delta E \Delta t \sim \hbar$ that the range of
reservoir energies that couple to each cavity mode goes to zero as
$t-t_0 \rightarrow \infty$.

\section{two sided cavity}
The generalization of the preceding to a two sided cavity,
$H_{SR},H_R \neq 0$, is straightforward since the reservoirs
couple independently to the cavity. For the right reservoir, we
define the input and output noise operators,
\begin{eqnarray}
G^{(in)}_n(t)=\sum_k\tilde{\kappa}_{n,k}b_k^{(in)}(t) \\
G^{(out)}_n(t)=\sum_k\tilde{\kappa}_{n,k}b_k^{(out)}(t)
\end{eqnarray}
where
\begin{eqnarray}
b_k^{(in)}(t)=b_k(t_0)e^{-i\omega_k (t-t_0)} \\
b_k^{(out)}(t)=b_k(t_1)e^{-i\omega_k (t-t_1)}
\end{eqnarray}
represent the input and output annihilation operators for the free
field modes of the right reservoir at time $t$. Associated with
the right reservoir noise operators are damping constants
\begin{equation}
\tilde{\gamma}_{n,m}=\pi \sum_k
\tilde{\kappa}_{n,k}\tilde{\kappa}^{*}_{m,k}\delta(\Omega_m-\omega_k),
\label{damp2}
\end{equation}
and radiative energy shifts
\begin{equation}
\tilde{\Delta}_{n,m}=P\sum_k
\tilde{\kappa}_{n,k}\tilde{\kappa}^{*}_{m,k}\frac{1}{\Omega_m-\omega_k}.
\end{equation}
The quantum Langevin equations for the cavity mode operators
expressed in terms of the input fields from the left and right are
then
\begin{widetext}
\begin{equation}
\dot{c}_n(t)=-i\Omega_n c_n(t)-\sum_m\left(
(\gamma_{n,m}+i\Delta_{n,m})c_m(t) +
(\tilde{\gamma}_{n,m}+i\tilde{\Delta}_{n,m})c_m(t) \right)
+F^{(in)}_n(t)+G^{(in)}_n(t) \label{lang_in2}
\end{equation}
\end{widetext}
One may also derive Langevin equations analogous to Eq.
(\ref{lang_out}) involving the output noise operators for the two
reservoirs, which can be used along with (\ref{lang_in2}) to
derive the boundary conditions for the noise operators in the left
and right reservoirs,
\begin{eqnarray}
G^{(in)}_n(t)-G^{(out)}_n(t)=2\sum_m \tilde{\gamma}_{n,m}c_m(t)
\label{bc2} \\
F^{(in)}_n(t)-F^{(out)}_n(t)=2\sum_m\gamma_{n,m}c_m(t).
\end{eqnarray}
Using the boundary condition $b_k(t_0)e^{i\omega_k
t_0}=-b_{-k}(t_1)e^{i\omega_k t_1}$ that corresponds to
(\ref{inoutbc}) for the free fields in the right reservoir, the
boundary condition for $b_k^{(in)}(t)$ and $b_k^{(out)}(t)$ can be
derived in the same manner as Eq. (\ref{bcmode}). One finds
\begin{eqnarray}
b_k^{(in)}(t)-b_{-k}^{(out)}(t)=2\pi\sum_m\tilde{\kappa}_{m,k}^*\delta(\Omega_m-\omega_k)c_m(t),
\label{bcmode2} \\
a_k^{(in)}(t)-a_{-k}^{(out)}(t)=2\pi\sum_m\kappa_{m,k}^*\delta(\Omega_m-\omega_k)c_m(t).
\label{bcmode3}
\end{eqnarray}

Equations (\ref{lang_in2}), (\ref{bcmode2}), and (\ref{bcmode3})
are the central result of this section. Given an initial state for
the two reservoirs at $t_0$, Eq. (\ref{lang_in2}) can be used to
calculate the state of the cavity at some later time. The mode
operators for the output field of the left and right reservoirs
can then be determined from the boundary conditions,
(\ref{bcmode2}) and (\ref{bcmode3}).

\section{output field statistics}
We illustrate how to utilize these results for a particular
initial state of the cavity plus reservoirs. Specifically, we
assume that the atom cavity initially contains no atoms and that
the right reservoir is likewise in the vacuum state. Furthermore,
the left reservoir contains a "beam" of fermions incident on the
barrier at $x=-a-d$, with a spatially uniform current density
equal to $\rho\hbar q/m$ where $\rho={\cal N}/L$ is the linear
atomic density and ${\cal N}$ is the total number of fermions.
This physical configuration is represented by the initial state
vector,
\begin{equation}
|\Psi (t_0)\rangle=\prod_{|k-q|\leq k_F}
a_k^{\dagger}(t_0)|0\rangle \label{initial}
\end{equation}
where $k_F=2\pi\rho$ is the Fermi momentum. $|\Psi (t_0)\rangle$
represents a zero temperature Fermi distribution that has been
given a Galilean boost that displaces the gas by $q$ in k-space.
This is analogous to the optical case where an incoherent white
light source is used to drive an optical cavity.

Even though $|\Psi (t_0)\rangle$ is a state with a fixed number of
atoms, it acts like a constant input flux of fermions on the
cavity. This is because of the implicit use of the Born
approximation in the derivation of the Langevin equations, i. e.
the reservoir is assumed to be so large that the back-action of
the system is negligible.

Using the results of the previous section we can calculate how the
state of the left and right reservoirs are modified due to their
coupling to the cavity. In particular, we focus on the occupation
numbers
\begin{eqnarray}
n^{(L)}_k(t)=a_k^{(out)\dagger}(t)a_k^{(out)}(t) \\
n^{(R)}_k(t)=b_k^{(out)\dagger}(t)b_k^{(out)}(t)
\end{eqnarray}
as well as the current density operators for the reservoirs
\[
J^{(L,R)}(x,t)=\sum_{\bar{q}}j^{(L,R)}_{\bar{q}}(t)e^{i\bar{q}x}
\]
where
\begin{eqnarray}
j^{(L)}_{\bar{q}}(t)=\frac{\hbar}{2mL}\sum_k(2k+\bar{q})a^{(out)\dagger}_k(t)a^{(out)}_{k+\bar{q}}(t)
\\
j^{(R)}_{\bar{q}}(t)=\frac{\hbar}{2mL}\sum_k(2k+\bar{q})b^{(out)\dagger}_k(t)b^{(out)}_{k+\bar{q}}(t)
\end{eqnarray}
are the spatial Fourier components of the current. Note that
$mL\langle j^{(L,R)}_{0}(t)\rangle $ is the average momentum in
the output fields.

Equation (\ref{lang_in2}) for $c_n(t)$ can be numerically
integrated but we note that because of the exponential dependence
of the reservoir-cavity coupling constants on the barrier height
and thickness, the off-diagonal coupling is much smaller than the
energy difference between the cavity modes,
\[
|\Omega_n-\Omega_m| \gg |\gamma_{n,m}+\tilde{\gamma}_{n,m}|,
 |\Delta_{n,m}+\tilde{\Delta}_{n,m}| \]
for $n \neq m$. In fact, a numerical evaluation of
$|\gamma_{n,m}+\tilde{\gamma}_{n,m}|$ and
$|\Delta_{n,m}+\tilde{\Delta}_{n,m}|$ using the coupling constants
of Appendix B indicate that these terms are at least two orders of
magnitude smaller than the energies of the cavity modes. It is
therefore an excellent approximation to neglect all off-diagonal
terms in the equations of motion. Furthermore, we can perform a
renormalization of the cavity mode energies by absorbing the
radiative energy shifts into them, $\Omega_m
+\Delta_{m,m}+\tilde{\Delta}_{m,m} \rightarrow \Omega_m$.

For times much longer than the lifetimes of the cavity modes,
$(t-t_0)(\gamma_{n,n}+\tilde{\gamma}_{n,n}) \gg 1$, the system
reaches a steady state with the solution,
\begin{equation}
c_m(t)=\sum_k \frac{\kappa_{m,k}a_k^{(in)}(t)+
\tilde{\kappa}_{m,k}b_k^{(in)}(t)}{i(\Omega_m-\omega_k)+\Gamma_m}
\label{ss}
\end{equation}
where $\Gamma_m=\gamma_{m,m}+\tilde{\gamma}_{m,m}$. It is easy to
see from Eq. (\ref{ss}) that the occupation numbers for the cavity
modes have a Lorentzian profiles,
\begin{equation}
\langle c^{\dagger}_m c_m\rangle =\sum_k \frac{|\kappa_{m,k}|^2
\langle n_k(t_0) \rangle}{(\Omega_m-\omega_k)^2+\Gamma_m^2}
\label{cavpop}
\end{equation}
where
\[
\langle n_k(t_0)\rangle =\Theta(k_F-|k-q|)
\]
are the occupation numbers for the input field. Due to the
broadband nature of the input field all cavity modes with energies
in the range $\omega_{k_F+q}-\omega_{k_F-q}$ will have a
significant population with higher energy cavity states having
larger populations due to the $|\kappa_{m,k}|$'s exponential
dependence on energy.

Using Eqs. (\ref{ss}) and the boundary conditions (\ref{bcmode2})
and (\ref{bcmode3}), we obtain steady state expectation values of
the occupation of the output field modes ,
\begin{widetext}
\begin{eqnarray}
\langle n^{(L)}_{-k} \rangle =& \left(1 -4\pi\sum_{m}
\frac{\Gamma_m \delta(\Omega_m-\omega_k) |\kappa_{m,k}|^2
}{(\Omega_m-\omega_k)^2+\Gamma_m^2} \right ) \langle
n_k(t_0)\rangle
+4\pi^2\sum_{m,n,k'}\frac{|\kappa_{m,k}|^2|\kappa_{m,k'}|^2
\delta(\Omega_m-\Omega_n)\delta(\Omega_m-\omega_k)}{(\Omega_m-\omega_{k'})^2+\Gamma_m^2}
\langle n_{k'}(t_0) \rangle \label{leftnum}
\end{eqnarray}
\begin{eqnarray}
\langle n^{(R)}_{k}
\rangle=&4\pi^2\sum_{m,n,k'}\frac{|\tilde{\kappa}_{m,k}|^2|\kappa_{m,k'}|^2
\delta(\Omega_m-\Omega_n)\delta(\Omega_m-\omega_k)}{(\Omega_m-\omega_{k'})^2+\Gamma_m^2}
\langle n_{k'}(t_0) \rangle . \label{rightnum}
\end{eqnarray}
\end{widetext}

Equations (\ref{leftnum}) and (\ref{rightnum}) represent the
changes in occupation numbers due to reflection and transmission
through the cavity. The most significant feature of Eq.
(\ref{leftnum}) is that fermions in the input beam are perfectly
reflected by the barrier, $\langle n^{(L)}_{-k} \rangle=\langle
n_k(t_0)\rangle$, unless there exists a cavity mode that is
degenerate in energy with state $k$ of the reservoir. The second
term represents the interference term between the fermions
reflected by the barrier and the fermions that have tunnelled into
the barrier and subsequently tunnelled back out into the left
reservoir. The last term in Eq. (\ref{leftnum}) represents the
tunnelling of atoms into mode $-k$ as a result of atoms from mode
$k'$ that have tunnelled into the cavity and then tunnel out of
the cavity.

In order to gain additional physical insight, we simplify these
expressions by noting that the denominator in Eq. (\ref{rightnum})
as well as the last term in Eq. (\ref{leftnum}) are sharply peaked
around $\omega_{k'}=\Omega_m=\omega_k$. Therefore we can replace
$k'$ with $k$ in this term and drop the summation over $k'$. Using
$\Gamma_m=2\gamma_{m,m}\approx 2\pi
|\kappa_{m,k}|^2\delta(\Omega_m-\omega_k)$ we have
\begin{eqnarray}
\langle n^{(L)}_{-k} \rangle \approx  \langle n_k(t_0) \rangle
-{\cal L} (\omega_k) \left ( \langle n_k(t_0) \rangle - \langle
n_{-k}(t_0) \rangle \right ), \label{resonant1} \\
\langle n^{(R)}_{k}\rangle \approx {\cal L} (\omega_k) \left (
\langle n_k(t_0) \rangle + \langle n_{-k}(t_0) \rangle \right ),
\label{resonant2}
\end{eqnarray}
where
\begin{equation}
{\cal L}(\omega_k)=\sum_m
\frac{\Gamma_m^2}{(\Omega_m-\omega_k)^2+\Gamma_m^2}. \label{D}
\end{equation}
When $\langle n_{-k}(t_0) \rangle =0$, Eq. (\ref{resonant1})
indicates that $\langle n^{(L)}_{-k} \rangle \approx 0$, a result
of the complete destructive interference between the fermions that
are directly reflected by the barrier and the fermions that tunnel
out of the cavity into the left reservoir. At the same time, Eq.
(\ref{resonant2}) indicates that fermions resonant with a cavity
mode tunnel through to the right side with unit probability,
$\langle n^{(R)}_{\pm k} \rangle \approx \langle n_{k}(t_0)
\rangle$. These transmission resonances are similar to the
situation in an optical Fabry-Perot cavity.

Fig. 3 shows a plot of $\langle n^{(L)}_{k} \rangle$ and $\langle
n^{(R)}_{k} \rangle$ using Eqs. (\ref{leftnum}) and
(\ref{rightnum}) for an incident beam with fermions occupying the
states $k=2\pi/L,...,(2\pi/L)\times 7501$. Note that we have taken
the reservoir to consist of discrete $k$ states, $k=2\pi n/L$,
with $n=0,\pm 1,...,\pm n_{max}$ and $n_{max}=L\sqrt{2mV_0}
/2\pi\hbar =10^4$. $V_0$ and $a$ were chosen so that the cavity
contained $50$ bound states. Each line in Fig. 3 corresponds to a
{\it single} reservoir $k$ state and as such, the width of the
lines are greatly exaggerated. The plots give good qualitative
agreement with the above discussion with each line located at the
$k$ state that is closest in energy to a particular cavity state.
The amplitude of the transmission resonances is close to $1$ for
the higher energy $k$ states, while for the lowest energy
resonances, the amplitudes are about $0.25$. The reduction in the
amplitude for the low energy states in comparison to
(\ref{resonant1}) and (\ref{resonant2}) comes from $\Gamma_m
\equiv 2\pi |\kappa_{m,k}|^2 \eta(\Omega_m) \gtrsim 2\pi
|\kappa_{m,k}|^2 \delta(\Omega_m-\omega_k)$ where $\eta
(\omega)\sim \omega^{-1/2}$ is the continuum density of states in
the reservoir (see Eq. (\ref{damp1})) .

\begin{figure}
\includegraphics[width=0.9\columnwidth]{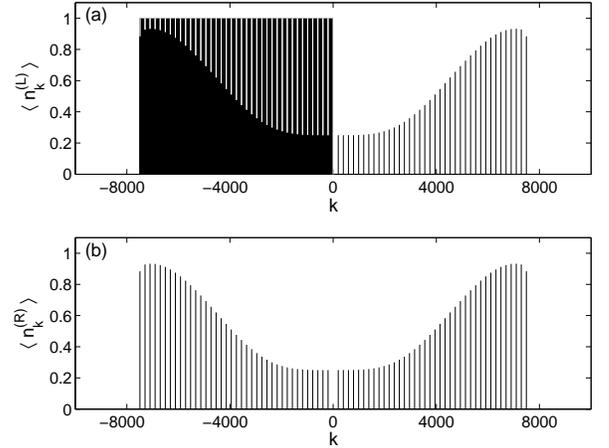}
\caption{ Plot of (a) $\langle n^{(L)}_{k}(t) \rangle$ and (b)
$\langle n^{(R)}_{k}(t) \rangle$ for $d/a=10^{-4}$ and
$a/L=1.2\times 10^{-5}$. $k$ is in units of $2\pi/L$.}
\end{figure}

The steady state current density in the output field of the left
and right reservoirs are given by,
\begin{widetext}
\begin{eqnarray}
\langle j_{-\bar{q}}^{(L)} \rangle =-\frac{\rho\hbar
q}{m}\delta_{\bar{q},0}-\frac{\hbar}{2mL}\sum_k(2k+\bar{q}) \left(
-2\pi \sum_m \kappa^*_{m,k+\bar{q}}\kappa_{m,k} \left(
\frac{\delta(\Omega_m-\omega_k)\langle n_{k+\bar{q}}(t_0)
\rangle}{-i(\Omega_m-\omega_{k+\bar{q}})+\Gamma_m} +
\frac{\delta(\Omega_m-\omega_{k+\bar{q}})\langle n_k(t_0)
\rangle}{i(\Omega_m-\omega_{k})+\Gamma_m} \right) \right. \nonumber \\
\left. +4\pi^2\sum_{m,n,k'}
\frac{\kappa_{m,k}\kappa_{n,k+\bar{q}}^*\kappa_{m,k'}^*\kappa_{n,k'}
\delta(\Omega_m-\omega_k)\delta(\Omega_n-\omega_{k+\bar{q}}) }
{(-i(\Omega_m-\omega_{k'})+\Gamma_m)(i(\Omega_n-\omega_{k'})+\Gamma_n)}
\langle n_{k'}(t_0) \rangle \right). \label{leftcurr}
\end{eqnarray}
and
\begin{eqnarray}
\langle j_{\bar{q}}^{(R)}\rangle
=\frac{\hbar}{2mL}\sum_k(2k+\bar{q}) \left( 4\pi^2\sum_{m,n,k'}
\frac{\tilde{\kappa}_{m,k}\tilde{\kappa}_{n,k+\bar{q}}^*\kappa_{m,k'}^*\kappa_{n,k'}
\delta(\Omega_m-\omega_k)\delta(\Omega_n-\omega_{k+\bar{q}}) }
{(-i(\Omega_m-\omega_{k'})+\Gamma_m)(i(\Omega_n-\omega_{k'})+\Gamma_n)}
\langle n_{k'}(t_0) \rangle \right), \label{rightcurr}
\end{eqnarray}
\end{widetext}
respectively.

The first term on the rhs of Eq. (\ref{leftcurr}) is the incident
current reflected by the barrier. The average momenta in the
output fields are proportional to $\langle j_0^{(L)}\rangle =
-\hbar/mL \sum_k k \langle n_{-k}^{(L)} \rangle$ and $\langle
j_0^{(R)} \rangle = \hbar/mL \sum_k k \langle n_{k}^{(R)}
\rangle$. In the left reservoir, $\langle j_0^{(L)} \rangle
\approx -\rho\hbar q/m $ since most of the fermions are perfectly
reflected by the barrier. In the right reservoir, the transmitted
current $\langle j_0^{(R)} \rangle \approx 0$ since  for those
states that are resonant with the cavity $\langle n_{k}^{(R)}
\rangle= \langle n_{-k}^{(R)}\rangle$ while for non-resonant
reservoir states, $\langle n_{k}^{(R)} \rangle \approx 0$.
Physically, this is due to the fact that atoms tunnel into the
right reservoir from a standing wave cavity mode and therefore,
they have equal probability to tunnel into states with positive
and negative momentum.

The $\bar{q}\neq 0$ terms are the spatial modulations in the
current density that build up in the reservoirs as a result of the
reservoir-cavity mode coupling. This can generate a coherence
between the $k$ and $k+\bar{q}$ modes when there is a finite
amplitude for an atom initially in state $k$ to tunnel into the
cavity and then tunnel back out of the cavity into state
$k+\bar{q}$. The change in $\langle j_{\bar{q}}^{(L)} \rangle$ is
of order $\kappa^2$ rather than $\kappa^4$ as was the case for Eq.
(\ref{leftnum}) since the current only involves generating a
coherence between $k$ and $k+\bar{q}$ rather than the transfer of
population. Furthermore, the $\kappa^2$ terms in (\ref{leftcurr})
are only finite for $|\omega_{k+\bar{q}}-\omega_k|<\Gamma_m$,
which implies that the coherence is only generated between
reservoir states whose energies lie within the linewidth of a
particular cavity mode. Consequently, decreasing the thickness and
height of the barriers will make the linewidths of the cavity
states larger and thereby increase the magnitude of $\bar{q}\neq
0$ components of the current.

Fig. 4 shows a plot of $\langle j_{\bar{q}}^{(L)} \rangle$ for
several values of $\bar{q}$. The $\bar{q}\neq 0$ components of the
current are about two orders of magnitude smaller than $\langle
j_{0}^{(L)} \rangle$ and they decay away with increasing
$\bar{q}$. We do not plot $\langle j_{\bar{q}}^{(R)} \rangle$
since the current is equal to $0$ to within our numerical
accuracy. This is because the cavity linewidths are so narrow that
it becomes nearly impossible to satisfy both the delta functions
in the numerator and the Lorentzian denominators of Eq.
(\ref{rightcurr}) for $\bar{q}\neq 0$ .

\begin{figure}
\includegraphics*[width=0.9\columnwidth]{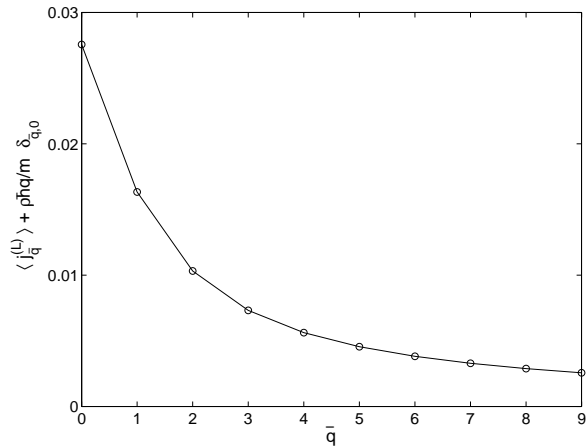}
\caption{ Plot of $\langle j_{\bar{q}}^{(L)} \rangle$ for the same
parameters as Fig. 3. $\bar{q}$ are in units of $2\pi/L$ and the
current is in units of $\hbar/2m a^2$. $\rho \hbar q/m=5.27$ in
these units.}
\end{figure}

\section{conclusion}
In this paper we have extended the quantum optical input-output
theory to atom cavities containing fermions. This formalism can
easily be applied to intra-cavity nonlinear atom optical processes
such as four-wave mixing between fermions \cite{moore2} or
coherent photoassociation of fermions into molecular bosons
\cite{timmermans}. In a future work we plan to extend this work to
treat non-Markovian dynamics \cite{savage} as well as including
the effects of the spatially delocalized single particle state
with energies greater than $V_0$.

\acknowledgments We thank J. V. Moloney for providing us with CPU
time on his parallel cluster. This work is supported in part by
the US Office of Naval Research under Contract No. 14-91-J1205, by
the National Science Foundation under Grants No. PHY00-98129, by
the US Army Research Office, by NASA Grant No. NAG8-1775, and by
the Joint Services Optics Program.

\section{Appendix A: Anticommutation relations between cavity and
noise operators} In this appendix, we examine some of the
consequences of Eqs. (\ref{lang_in}) and (\ref{bc}). Correlations
between the input and output fields and the cavity modes may be
expressed in terms of the anticommutators for the cavity and noise
operators, i. e. nonvanishing equal time anticommutators imply
that the operators are not independent. From Eq. (\ref{lang_in}),
one sees that the solution for $c_n(t)$ can be expressed in terms
of the initial conditions for the cavity operators, $c_n(t_0)$,
and the $a_k(t_0)$. It follows immediately that
$\{c_n(t),F_m^{(in)}(t')\}=0$ for all $t$ and $t'$ since the only
non-vanishing anti-commutators at $t_0$ are between creation and
annihilation operators. In a similar manner, it immediately
follows from Eq. (\ref{lang_out}) that
$\{c_n(t),F_m^{(out)}(t')\}=0$. Formally integrating Eq.
(\ref{lang_in}) from $t_0$ to $t$ shows that $c_n(t)$ depends on
$F_m^{(in)}(t')$ for $t'<t$ (note that $c_n(t)$ will, in general,
depend on all of the noise operators, $F_m^{(in)}(t')$, not just
$n=m$, due to the intra-cavity mode coupling). Using Eq.
(\ref{noisecomm}) it then follows that
\begin{equation}
\{c_n(t-\tau),F_m^{(in)\dagger}(t)\}=0 \label{cin1}
\end{equation}
for $\tau>0$. This is nothing more than a statement of causality,
the system can only depend on the past input. In a similar manner,
formal integration of Eq. (\ref{lang_out}) from $t+\tau$ to $t_1$
implies that
\begin{equation}
\{c_n(t+\tau),F_m^{(out)\dagger}(t)\}=0 \label{cout1}
\end{equation}
for $\tau>0$. Again, this is nothing more than the statement that
the output of the system can only depend on the state of the
system in the past. Using Eqs. (\ref{cout1}) and (\ref{bc}), one
obtains
\[
\{c_n(t+\tau),F_m^{(in)\dagger}(t)\}=2\sum_p\gamma_{m,p}^{*}\{c_n(t+\tau),c_p^{\dagger}(t)\}
\]
while using Eq. (\ref{cin1}) along with (\ref{bc}) gives,
\[
\{c_n(t-\tau),F_m^{(out)\dagger}(t)\}=-2\sum_p\gamma_{m,p}^{*}\{c_n(t-\tau),c_p^{\dagger}(t)\}
\]
for $\tau>0$. It follows from Eq. (\ref{bc}) that the equal time
anticommutators can be written as
$\{c_n(t),F^{(in/out)\dagger}_m(t)\}=\pm
\sum_p\gamma^{*}_{m,p}\{c_n(t),c^{\dagger}_p(t)\}+A_{n,m}$. The
operator $A_{n,m}$ is determined from the equations of motion,
(\ref{lang_in}) and (\ref{lang_out}), to be $-i\sum_p
\Delta^{*}_{m,p}\{c_n(t),c^{\dagger}_p(t)\}$. If we define the
step function $u(\tau)$ as
\[
u(\tau)=
  \begin{cases}
   1 &, \quad \tau>0 \\
   1/2 &, \quad \tau=0 \\
   0 &, \quad \tau<0
  \end{cases}
\]
then the anticommutators for arbitrary $\tau$ are given by
\begin{widetext}
\begin{eqnarray}
\{c_n(t+\tau),F_m^{(in)\dagger}(t)\}=2u(\tau)\sum_p(\gamma_{m,p}^{*}-i\Delta^{*}_{m,p}\delta_{\tau,0})\{c_n(t+\tau),c_p^{\dagger}(t)\}
\label{in_anti}
\\
\{c_n(t+\tau),F_m^{(out)\dagger}(t)\}=-2u(-\tau)\sum_p(\gamma_{m,p}^{*}+i\Delta^{*}_{m,p}\delta_{\tau,0})\{c_n(t+\tau),c_p^{\dagger}(t)\}.
\end{eqnarray}
\end{widetext}

Finally, we show that Eq. (\ref{lang_in}) preserve the equal time
anticommutators for the system operators. Since (\ref{lang_in})
constitute an $N\times N$ system of equations, an explicit
solution will be non-trivial for $N>2$. However, it is already
apparent that $\{c_n(t),c_m(t)\}=0$ since, as we stated before,
$c_n(t)$ can be expressed in terms of the initial states for the
operators at $t_0$ and
$\{c_n(t_0),c_m(t_0)\}=\{c_n(t_0),a_k(t_0)\}=0$. We are therefore
left with showing that $\{c_n(t),c_m^{\dagger}(t)\}=\delta_{n,m}$.
Calculating the derivative of the anticommutator using
(\ref{lang_in}) and (\ref{in_anti}), one obtains,
\[
\frac{d}{dt}
\{c_n(t),c_m^{\dagger}(t)\}=-i(\Omega_n-\Omega_m)\{c_n(t),c_m^{\dagger}(t)\}.
\]
Integrating this expression and using the initial condition,
$\{c_n(t_0),c^{\dagger}_m(t_0)\}=\delta_{n,m}$, we obtain the
desired result.

\section{Appendix B: Tunnelling coupling constants}
In this appendix we derive explicit expressions for the tunnelling
coupling constants $\kappa_{n,k}$ and $\tilde{\kappa}_{n,k}$.
Tunnelling in many-body systems was first treated by Bardeen
\cite{bardeen} and later elaborated on by Prange \cite{prange} and
Harrison \cite{harrison} in the context of electron tunnelling
across insulator junctions. Bardeen showed that if there exists a
potential barrier separating two regions of space, then the
tunnelling matrix element, $T_{a,b}$, for the tunnelling of a
particle from state $\phi_b(x)$ on one side of the barrier into
the state $\phi_a(x)$ on the opposite side of the barrier is given
by the off-diagonal current density in the barrier,
\begin{equation}
T_{a,b}=\frac{-\hbar^2}{2m}\left[ \phi_a^*(x)\frac{d}{dx}\phi_b(x)
-\phi_b(x)\frac{d}{dx}\phi_a^*(x) \right]_{x=x_1}. \label{TT}
\end{equation}
Here $x_1$ is a point inside the barrier and $\phi_a(x)$ and
$\phi_b(x)$ are the continuations of the single particle wave
functions into the barrier where they decay exponentially.
$T_{a,b}$ is independent of the choice of $x_1$ provided the
energy difference between the two states $\phi_a$ and $\phi_b$ is
much less than the height of the barrier. The relationship between
(\ref{TT}) and the overlap of the Hamiltonian between states
localized on either side of the barrier is discussed in
\cite{prange}.

For the system illustrated in Fig. 1, we identify
$-i\hbar\kappa^*_{n,k}$ with $T_{a,b}$ when
$\phi_a(x)=\phi^{(l)}_k(x)$ and $\phi_b(x)=\phi^{(s)}_n(x)$.
Similarly, $-i\hbar\tilde{\kappa}^*_{n,k}$ is equal to $T_{a,b}$
when $\phi_a(x)=\phi^{(r)}_k(x)$ and $\phi_b(x)=\phi^{(s)}_n(x)$.

For the purpose of calculating the coupling constants, we take the
$\phi^{(s)}_n(x)$ to be the eigenstates of the finite potential
well corresponding to $d\rightarrow \infty$. In the classically
allowed region, $-a\leq x \leq a$, $\phi^{(s)}_n(x)$ is
proportional to $\cos (K_nx)$ for $n$ even and $\sin (K_nx)$ for
$n$ odd while inside the barrier $\phi^{(s)}_n(x)\sim
\exp(-\sqrt{2m(V_0-\hbar\Omega_n)/\hbar^2}|x|)$. The cavity wave
numbers $K_n$ are determined by the solutions to
\begin{equation}
K_na \tan K_na=\sqrt{\beta^2-(K_na)^2}.
\end{equation}
for $n$ even and
\begin{equation}
K_na \cot K_na=-\sqrt{\beta^2-(K_na)^2}.
\end{equation}
for $n$ odd \cite{merzbacher}.

For even $n$ we find,
\begin{equation}
\kappa_{n,k}=\frac{-i\hbar\sigma}{m}\sqrt{\frac{2K_n}{L(1+(\sigma/k)^2)(K_na+\cot
K_na)}}e^{-\sigma d}\cos K_na
\end{equation}
with
\begin{equation}
\tilde{\kappa}_{n,k}=\kappa_{n,k} \label{coupleeven}
\end{equation}
In a similar manner, we find for $n$ odd,
\begin{equation}
\kappa_{n,k}=\frac{i\hbar\sigma}{m}\sqrt{\frac{2K_n}{L(1+(\sigma/k)^2)(K_na-\tan
K_na)}}e^{-\sigma d}\sin K_na
\end{equation}
with
\begin{equation}
\tilde{\kappa}_{n,k}=-\kappa_{n,k}. \label{coupleodd}
\end{equation}

In both cases, $\sigma=\sqrt{2mV_0/\hbar^2-k^2}$ is the inverse of
the penetration depth of the reservoir state into the barrier and
$\beta^2=2ma^2V_0/\hbar^2$ is the dimensionless barrier height.

An interesting consequence of Eqs. (\ref{coupleeven}) and
(\ref{coupleodd}) is that if $n$ is even and $m$ is odd or vice
versa, then
\begin{equation}
(\gamma_{n,m}+i\Delta_{n,m})+
(\tilde{\gamma}_{n,m}+i\tilde{\Delta}_{n,m})\equiv 0.
\end{equation}
On the other hand if $n$ and $m$ are both even or both odd, then
\begin{equation}
(\gamma_{n,m}+i\Delta_{n,m})+
(\tilde{\gamma}_{n,m}+i\tilde{\Delta}_{n,m})=2(\gamma_{n,m}+i\Delta_{n,m}).
\end{equation}
Since $\phi_n^{(s)}(x)$ is an eigenstate of the parity operator
with parity $(-1)^n$, it follows from Eq. (\ref{lang_in2}) that
for a two-sided cavity only states of the same parity are coupled.
This is a direct consequence of the our model system in Fig. 1
being invariant under spatial reflections, which implies that even
when the coupling of the cavity states to the reservoirs is taken
into account, parity will still be a good quantum number for the
cavity states. No such result holds for the single-sided cavity
since the cavity plus reservoir system is no longer invariant with
respect to reflections.

Finally we note that the approximation $\omega_k \approx \Omega_n$
used in deriving $\kappa_{n,k}$ becomes exact for the evaluation
of the $\gamma_{n,n}$ and the cavity boundary conditions
(\ref{bcmode}) and (\ref{bcmode2}) since in these expression
$\kappa_{n,k}$ is always multiplied by $\delta
(\Omega_n-\omega_k)$.


\begin{references}
\bibitem{schrodrep} The momentum operator is defined in the
Schr\"{o}dinger representation. At the initial time $t_0$ the
operators in the Heisenberg and Schr\"{o}dinger representations
are equal.
\bibitem{manipnote} $D(x)F^{(out)}(t)D^{\dagger}(x)$ can be evaluated
using $\hat{a}_{-k}(t_1)=-\hat{a}_k(t_0)-\sum_m
\kappa^{*}_{m,k}\int_{t_0}^{t_1}dt'e^{i(\omega_k-\Omega_m)t'}\hat{c}_m(t')$
\end{references}
\end{document}